\documentclass[%
 reprint,
 amsmath,amssymb,
 aps,
]{revtex4-2}

\usepackage{graphicx}% Include figure files
\usepackage{dcolumn}% Align table columns on decimal point
\usepackage{bm}% bold math
\newcommand{\1}{{\mathbb 1}}

\usepackage{amssymb}
\usepackage{epsfig}
\usepackage{color}
\usepackage{amsfonts}
\usepackage{amsmath}
\usepackage{bbold}

\newcommand{\spacer}{\rule[0cm]{0cm}{0cm}}

\newtheorem{theorem}{Theorem}
\newtheorem{lemma}{Lemma}
\newtheorem{proposition}{Proposition}
\newtheorem{corollary}{Corollary}

\newcounter{statementnumber}
\renewcommand{\thestatementnumber}{\arabic{section}.\arabic{statementnumber}}

\def\mb{$\begin{displaystyle}}
\def\me{\end{displaystyle}$\ }

\DeclareMathOperator{\wt}{wt}

\DeclareMathOperator{\GSym}{GSym}

\newcommand{\SP}{SP}
\newcommand{\CP}{CP}

\newcommand{\ket}[1]{\left| #1 \right\rangle}
\newcommand{\bra}[1]{\left\langle #1 \right|}

\newcommand{\Stab}{\mbox{\rm Stab}}
\newcommand{\R}{{\mathbb R}}

\newcommand{\C}{{\mathbb C}}

\begin{document}
\title{Local unitary classes of states invariant under permutation subgroups}

\author{David W. Lyons}
\email{lyons@lvc.edu}
\affiliation{Mathematical Sciences, Lebanon Valley College, PA, USA}
\author{Jesse R. Arnold}
\email{jra005@lvc.edu}
\affiliation{Mathematical Sciences, Lebanon Valley College, PA, USA}
\author{Ashley F. Swogger}
\email{afs004@lvc.edu}
\affiliation{Mathematical Sciences, Lebanon Valley College, PA, USA}

%\date{\today}% It is always \today, today,
             %  but any date may be explicitly specified
\date{22 September 2021, accepted 7 March 2022, published 25 March 2022}

\begin{abstract}
  The study of entanglement properties of multiqubit states that are
  invariant under permutations of qubits is motivated by potential
  applications in quantum computing, quantum communication, and quantum
  metrology. In this work, we generalize the notions of symmetrization,
  Dicke states, and the Majorana representation to the alternating,
  cyclic, and dihedral subgroups of the full group of permutations. We
  use these tools to characterize states that are invariant under these
  subgroups and analyze their entanglement properties.
\end{abstract}

%\keywords{Suggested keywords}%Use showkeys class option if keyword
                              %display desired

%\tableofcontents
\maketitle

%\tableofcontents

\section{Introduction}

Entangled states of many quantum bits are essential resources in
emerging technologies including quantum computing, secure
communication, and measurement devices that promise to outperform
`classical' digital technologies in fundamental ways. While such
applications drive the study of multiparticle entanglement, a deeper
motivation is to achieve insight in the foundations of physics.

In its full generality, entanglement is a hard problem: it is not
reasonable to expect a full classification of
multiparticle entanglement types~\cite{meyer01p}. A more modest, yet
still valuable goal is to identify and classify families of entanglement
types that are useful in protocols in quantum computation and communication.

One such family is the symmetric states, that is, states of composite
systems that are invariant under permutations of the
subsystems. Fruitful studies of permutation invariant states where the
general case remains intractable include: geometric measure of
entanglement~\cite{aulbach2010,aulbach2010b,markham2010}, efficient
tomography~\cite{toth2010}, classification of states equivalent under
stochastic local operations and classical communication
(SLOCC)~\cite{bastin2009,bastin2010}, and our own work on classification
of states equivalent under local unitary (LU)
transformations~\cite{symmstatespaper,symmstates2}. Recent work~\cite{burchardt2021} of
Burchardt et al., and also in this paper, generalizes the study of entanglement properties
of permutationally invariant states to states that are invariant under
the action of {\em subgroups} of the full permutation group. 

The investigation in this paper is motivated by the following
example. Higuchi and Sudbery~\cite{higuchi00} identified the
state 
\begin{align}\label{M4def}
  \ket{M_4} =\frac{1}{\sqrt{6}}[&{(\ket{0011} + \ket{1100})}\\
    \nonumber
     + \omega &{(\ket{1010} + \ket{0101})}\\ \nonumber
    + \omega^2 &{(\ket{1001} + \ket{0110})}]
\end{align}
(where $\omega = e^{2 \pi i/3}$) in a study seeking to analyze various
maximal entanglement properties. The state $\ket{M_4}$ has the property
that it has the maximum average two-qubit bipartite entanglement,
averaged over all partitions into 2-qubit
subsystems. We found~\cite{maxstabnonprod2} that the
  state $\ket{M_4}$ and its conjugate $\ket{\overline{M}_4}$ are
  characterized, up to local unitary equivalence, by their
  invariance under the action of local unitary operators of
  the form $U^{\otimes 4}$, for arbitrary 1-qubit unitaries $U$. An
  application of this fact is a code, using
  $\ket{M_4},\ket{\overline{M}_4}$ for logical qubits
  $\ket{0_L},\ket{1_L}$, that is unaffected by noise that takes the form
  of the same unitary evolution on each qubit. In addition to {\em
    local} unitary invariance properties, $\ket{M_4}$ has {\em nonlocal} permutation
  invariance under the subgroup $A_4$ of even permutations, that is,
  permutations that are products of an even number of transpositions of
  subsystems. The state $\ket{M_4}$ is {\em not} invariant under the
  full group of permutations of qubits: any odd number or transpositions
  of subsystems takes $\ket{M_4}$ to $\ket{\overline{M}_4}$, and
  vice versa.

{Together, these observations about maximum
  entanglement properties, local unitary invariance, nonlocal permutation
  invariance, and potential application to quantum information
  protocols, motivate the study of entanglement properties and potential
  applications of states that are invariant under subgroups of the full
  permutation group. We give a complete local unitary classification of states that are
invariant under the alternating groups $A_n$ in
Section~\ref{ansection}, and a partial classification for the cyclic groups $C_n$ and
the dihedral groups $D_n$ in Section~\ref{cndnsection}. We present
evidence for applications in Section~\ref{applicationssection}. We begin, in
Section~\ref{prelimsect}, by establishing basic definitions and tools
for the study of $G$-invariance, including generalizations of the
$S_n$-invariant Dicke states and symmetrization constructions. Proofs of
most Lemmas, Propositions, and Theorems are in the Appendix; a few short
proofs are in the main body of the paper.}

\section{Permutation subgroup invariance}\label{prelimsect}

\subsection{Invariance up to phase}

Let $G$ be a subgroup of the permutation group $S_n$. A permutation
$\sigma$ in $G$ acts
on the the Hilbert space $(\C^2)^{\otimes n}$ of $n$-qubit states by
\begin{align}\label{snaction}
  &\sigma (\ket{\phi_1}\otimes \ket{\phi_2}\otimes \cdots
  \otimes \ket{\phi_n})\\
  \nonumber
  &=
  \ket{\phi_{\sigma^{-1}(1)}}\otimes \ket{\phi_{\sigma^{-1}(2)}}\otimes \cdots
  \otimes \ket{\phi_{\sigma^{-1}(n)}}
\end{align}
where the $\ket{\phi_k}$ are 1-qubit states. The effect of
$\sigma$ is to move the entry in position $j$ to the position $\sigma(j)$.
We say that an $n$-qubit state $\ket{\psi}$ is $G$-invariant {\em up to phase}
if, for all $\sigma\in
G$, there exists a nonzero scalar $t_\sigma$ such that 
\begin{equation}
  \label{ginvdefstate}
\sigma\ket{\psi}=t_\sigma \ket{\psi}.
\end{equation}
Invariance up to phase is a generalization of belonging to the {\em
  symmetric subspace}, that is, having the property that $t_\sigma=1$
for all $\sigma$.  The state $\ket{M_4}$ (see~(\ref{M4def})
above) is invariant up to phase, but $\ket{M_4}$ is not in the $A_4$-symmetric subspace. For example, we have
$(132)\ket{M_4}=\omega^2 \ket{M_4}$, where $\omega = e^{2\pi i/2}$ and $(132)$ denotes the
3-cycle permutation $1\to 3\to 2\to 1$. Table~\ref{tab:a4m4actvalues}
shows a complete list of values $t_\sigma$. {\em [Notation convention: In
Table~\ref{tab:a4m4actvalues}, and throughout this paper, we use
standard cycle notation to denote permutations. For distinct integers
$a_1,a_2,\ldots,a_k$ in the range $1\leq a_i\leq n$, the symbols
$(a_1a_2\ldots a_k)$ denote the cyclic permutation
$$ a_1\to a_2\to a_3 \to \cdots \to a_n \to a_1.
$$
Products of cycles $\kappa_1\kappa_2\ldots \kappa_t$ are read from right
to left as function compositions.]}

\begin{table}[h!]
  \centering
  %% \begin{tabular}{c|cccccccccccc}
  %%   $\sigma$ & $(12)(34)$ & $(13)(24)$ & $(14)(23)$ & $(123)$ & $(132)$ &
  %%   $(124)$ & $(142)$ & $(234)$ & $(243)$ & $(134)$ & $(143)$ &
  %%   $e$\\ \hline
  %%   $t_\sigma$ & $1$ & $1$ & $1$ & $\omega$ & $\omega^2$ & $\omega^2$ &
  %%   $\omega$ & $\omega^2$ & $\omega$ & $\omega$ & $\omega^2$ & $1$       
  %% \end{tabular}
  \begin{tabular}[t]{c|c}
    $\sigma$ & $t_\sigma$ \\ \hline
    $e$        & $1$        \\
  \end{tabular}\hspace*{.25in}
  \begin{tabular}[t]{c|c}
    $\sigma$ & $t_\sigma$ \\ \hline
    $(12)(34)$ & $1$        \\
    $(13)(24)$ & $1$        \\
    $(14)(23)$ & $1$        \\
  \end{tabular}\hspace*{.25in}
  \begin{tabular}[t]{c|c}
    $\sigma$ & $t_\sigma$ \\ \hline
    $(123)$    & $\omega$   \\ 
    $(132)$    & $\omega^2$ \\
    $(124)$    & $\omega^2$ \\
    $(142)$    & $\omega$   \\
    $(234)$    & $\omega^2$ \\
    $(243)$    & $\omega$   \\
    $(134)$    & $\omega$   \\
    $(143)$    & $\omega^2$ \\
  \end{tabular}
  \caption{Values of $t_\sigma$ for $\sigma \in A_4$ acting on $\ket{M_4}$, where
  $\omega=e^{2\pi i/3}$}
  \label{tab:a4m4actvalues}
\end{table}

Because equation~(\ref{snaction}) defines a group action of $G$ on
$n$-qubit space, the function $t\colon G\to U(1)$ in
equation~(\ref{ginvdefstate}) has the property that
$t_{\sigma\tau}=t_\sigma t_\tau$ for all $\sigma,\tau$ in
$G$. In other words, $t$ is a homomorphism of groups. From this we
obtain useful properties such as the following.
\begin{itemize}
\item $t_e=1$, where $e\in G$ is the identity permutation
\item $t_{\sigma^{-1}}=t_\sigma^{-1}$ for $\sigma \in G$
  \item if $\sigma^m=1$ then $t_\sigma^m=1$
\end{itemize}
We record this key observation as a Proposition.

\begin{proposition}\label{tishomism}
  Let $G$ be a subgroup of $S_n$, let $\ket{\psi}$ be a
  $G$-invariant state, and let $t\colon G\to U(1)$ be given by
  $\sigma\ket{\psi}=t_\sigma\ket{\psi}$. Then $t$ is a group
  homomorphism. That is, we have $t_{\sigma\tau}=t_\sigma t_\tau$ for
  all $\sigma,\tau$ in $G$.
\end{proposition}

%% {\bf Proof.} The statement follows directly from the fact that the
%% action of $G$ on $n$-qubit space is a group action, that is, we have
%% $$t_{\sigma\tau}=t_\sigma t_\tau
%% $$
%% for all $\sigma,\tau$ in $G$.

In mathematical terminology, the homomorphism in
Proposition~\ref{tishomism} is an element of the {\em dual group} of
$G$~\cite{dummitandfoote}. In place of saying ``dual group element'', we
will use the unofficial, but more descriptive term {\em phase
  homomorphism} to refer to a map $G\to U(1)$ arising from a
$G$-invariant state $\ket{\psi}$ by the equations $\sigma\ket{\psi} =
t_\sigma\ket{\psi}$ for $\sigma\in G$.

We conclude this subsection with a remark about invariance up to phase
for $G=S_n$. An example of a permutationally invariant state that is not
in the symmetric subspace is
$\ket{s} = \frac{1}{\sqrt{2}}(\ket{01}-\ket{10})$, for which we have
$(12)\ket{s}=-\ket{s}$. The following proposition shows that
$\ket{s}$ is the {\em only} example for $S_n$-invariance up to phase 
is different from belonging to the symmetric subspace. The proof is in the Appendix.

\begin{proposition}\label{singletonlyphaseinv}
  Let $\ket{\psi}$ be an $n$-qubit state for $n\neq
  2$. Suppose that $\sigma\ket{\psi} = t_\sigma\ket{\psi}$ for every
  $\sigma$ in $S_n$, where the $S_n$ action on $n$-qubit vector space is
  given by~(\ref{snaction}). Then $t_\sigma=1$ for all $\sigma$.
\end{proposition}

\subsection{Generalized Dicke forms}

In this subsection we develop a key tool for the analysis of
$G$-invariant states that generalizes the Dicke states for
$S_n$-invariant states.

Let $G$ be a subgroup of $S_n$. The group $G$ acts on $n$-bit strings by
\begin{equation}\label{permactonstring}
  g(i_1i_2\ldots i_n) = i_{g^{-1}(1)}i_{g^{-1}(2)}\ldots i_{g^{-1}(n)}
\end{equation}
for $g\in G$. We will write $[I]$ to denote the $G$-orbit
$$ [I] = \{gI\colon g\in G\}
$$
and we will write $\Stab^G_I$ to denote the stabilizer subgroup
$$\Stab^G_I = \{g\in G\colon gI=I\}
$$
of the bit string $I$.

Let $\ket{\psi}$ be a $G$-invariant state (throughout this paper, we
adopt~(\ref{ginvdefstate}) for the definition of $G$-invariant state,
and will omit the phrase ``up to phase''). Let $\ket{\psi}=\sum_I c_I
\ket{I}$ be the expansion of $\ket{\psi}$ in the computational
basis. For $g\in G$, we have
\begin{equation}\label{gglobalphase}
  g\ket{\psi} = t_g\ket{\psi} = \sum_I t_gc_I \ket{I}
\end{equation}
and we also have
\begin{equation}\label{glocalperm}
  g\ket{\psi} = \sum_I c_I \ket{gI}.
\end{equation}
Comparing the $\ket{J} = \ket{gI}$ term in equations~(\ref{gglobalphase})
and~(\ref{glocalperm}) above, we have
\begin{equation}\label{gitoJconstt}
  c_I = t_gc_{J}  \mbox{ for all $g\in G$ such that $gI=J$}.
\end{equation}
It follows that if $c_I\neq 0$ and $gI=I$, then $t_g=1$. Thus we have
the following.

{\bf Observation.} If $\ket{\psi}$ is $G$-invariant, with phase homomorphism
$t\colon G\to U(1)$ given by $g\ket{\psi}=t_g\ket{\psi}$ for all $g\in
G$, then $t$ is constant on $\Stab^G_I$ for all $I$ such that $c_I\neq
0$.

This observation leads to the following representations of the
$G$-invariant state $\ket{\psi}$ that generalize the Dicke form for
$S_n$-invariant states. For each $G$-orbit $[I]$, choose a fixed
representative $L_{[I]}$. (For example, we could choose $L_{[I]}$ to be
the bit string that represents the largest binary integer among all the
elements of $[I]$.)
We will write $c_{[I]}$ to denote $c_{L_{[I]}}$. We have the following
{\em generalized Dicke forms} for $\ket{\psi}$.
\begin{align}
  \ket{\psi} &= \sum_{[I]} \frac{c_{[I]}}{\sqrt{|\Stab^G_I|}}
  \sum_{g\in G} t_g^{-1} \ket{gI}\\ \label{dickeform}
  &= \sum_{[I]} c_{[I]} \sum_{J\in [I]} t_g^{-1}\ket{J} \hspace*{.5in}\mbox{(where $gL_{[I]}=J$)}
\end{align}
That the expression $t_g$ in~(\ref{dickeform}) is independent of the choice
of $g$ (as long as $gL_{[I]}=J$) is justified by~(\ref{gitoJconstt}).

We have shown that any $G$-invariant state has a Dicke form. Now we show
the converse (the proof is in the Appendix).

\begin{proposition}\label{dickeformisginv}
Let $G$ be a subgroup of $S_n$, let $c_{[I]}$ be
complex constants, one for each $G$-orbit on $n$-bit strings, and let $t\colon G\to
U(1)$ be a homomorphism that is constant on any stabilizer $\Stab^G_I$
for which $c_I\neq 0$. The state
$$  \ket{\psi}= \sum_{[I]} c_{[I]} \sum_{J\in [I]} t_g^{-1}\ket{J} \hspace*{.5in}\mbox{(where $gL_{[I]}=J$)}
$$
is $G$-invariant, and satisfies $g\ket{\psi} = t_g\ket{\psi}$ for all
$g\in G$.
\end{proposition}

We note here that the Dicke form~(\ref{dickeform}) is indeed a
generalization of the decomposition $\ket{\psi} = \sum_{w=0}^n d_w
\ket{D^w_n}$ of a state $\ket{\psi}$ in the ($S_n$-)symmetric subspace,
where
\begin{equation}\label{sndicke}
\ket{D^w_n} = \frac{1}{\sqrt{n\choose w}}
\sum_{I\colon \wt I = w}\ket{I}
\end{equation}
is the weight $w$ Dicke state. (Here,
the symbols ``$\wt I$'' denote the (Hamming) weight of the binary string
$I$, that is, the number of 1's in $I$).

We conclude this subsection with a definition of the {\em generalized
  Dicke states}. Given
a homomorphism $t\colon G\to U(1)$ and a $G$-orbit $[I]$ such that $t$ is
constant on $\Stab^G_I$, we refer to the
states
\begin{align}\label{unnormgendicke}
  \ket{\widetilde{D}_t^{[I]}} &= \sum_{J\in [I]} t_g^{-1}
 \ket{J} \hspace*{.5in}\mbox{(where $gL_{[I]}=J$)}\\
  \ket{D_t^{[I]}} &= \frac{1}{\sqrt{|[I]|}} \ket{\widetilde{D}^{[I]}}
\end{align}
as the unnormalized (respectively, normalized) generalized Dicke state
for the $G$-orbit $[I]$ with respect to the homomorphism $t$. 

\subsection{The permutation subgroups $A_n$, $C_n$, and $D_n$}

  \begin{figure}
  \centering{
  \includegraphics[scale=.50]{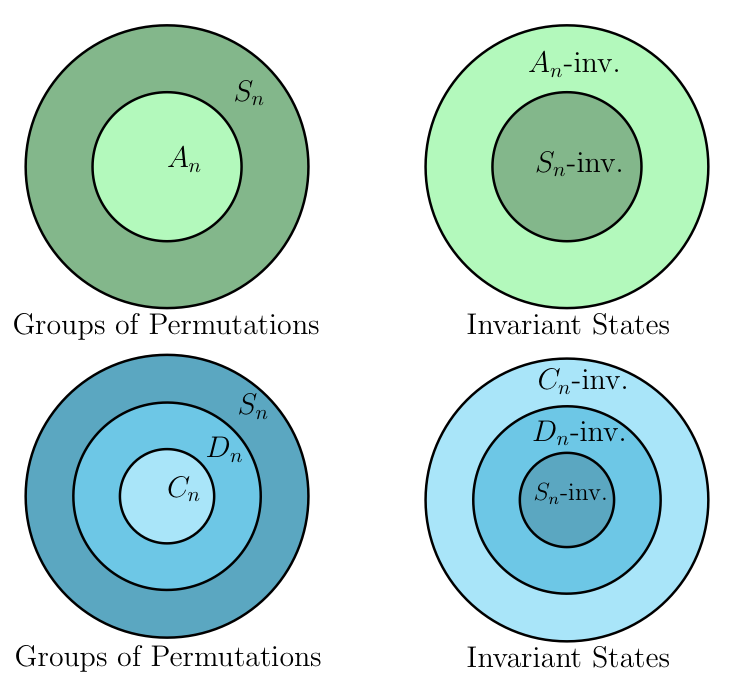}
  \caption{Subgroups of $S_n$ and corresponding sets of invariant states}
  \label{gpsandstatesfig}
  }
\end{figure}

In this paper, we consider the following permutation subgroups: the
alternating group $A_n$, the cyclic group $C_n$, and the dihedral group
$D_n$, defined as follows.
\begin{itemize}
\item [$A_n$:] the alternating group is the set of permutations that can
  be written as a product of an even number of permutations
  \item [$C_n$:] the cyclic group is the group generated by the full cycle
    $\epsilon=(12\cdots n)$
    \item [$D_n$:] the dihedral group is the group generated by the full
      cycle $\epsilon=(12\cdots n)$ and the ``mirror reflection'' $\tau
      = \prod_{j=1}^{\lfloor \frac{n+1}{2}\rfloor} (j,n+1-j)$
\end{itemize}
In terms of the action~(\ref{permactonstring}) of permutations on bit
strings, the effect of $\tau$ is string reversal, that is, we have
$$\tau(i_1i_2\cdots i_{n-1}i_n) = (i_ni_{n-1}\cdots i_2i_1).
$$
In what follows, we will use the facts that $A_n$ is generated by
3-cycles~\cite{gallian2017contemporary}, 
% Gallian, Ch 5 Ex 57
that $C_n$ is generated by $\epsilon$
(by definition), and that $D_n$ is generated by $\epsilon$ and $\tau$.
\begin{align*}
  A_n &= \left\langle (ijk) \right\rangle \hspace*{.2in}(1\leq i,j,k\leq n,\; i,j,k
    \mbox{ distinct})\\
  C_n &= \left\langle \epsilon \right\rangle\\
  D_n &= \left\langle \epsilon,\tau \right\rangle
\end{align*}
Figure~\ref{gpsandstatesfig} illustrates the inclusions among the 
groups $S_n$, $A_n$, $D_n$, and $C_n$, and the reversed inclusions among the corresponding invariant
states. In the sections that follow, we focus on the characterization of
states that lie in the annular rings of
Figure~\ref{gpsandstatesfig}, that is, we characterize states that are
$A_n$-invariant and are {\em not} fully $S_n$ invariant, and states that are
$C_n$-invariant and are {\em not} $D_n$-invariant.
Table~\ref{dickesummary} shows a summary table of the (unnormalized)
generalized Dicke states~(\ref{unnormgendicke}) for $G=S_n,C_n,D_n$ (the
case of $A_n$ is treated in Subsection~\ref{andickesubsection}).

  \begin{center}
\begin{table}
  \centering{
    \begin{tabular}{c|c}
      $G$  & $\displaystyle \ket{\widetilde{D}_t^{[I]}}$\\ \hline
      $S_n$   & $\displaystyle \sum_{I\colon \wt I=w}\ket{I}$\\
      $C_n$ & $\displaystyle \sum_{k=0}^{n-1}
      t_\epsilon^{-k} \ket{\epsilon^k I}$\\
      $D_n$  & $\displaystyle \sum_{a=0}^1\sum_{k=0}^{n-1}
      t_\tau^a t_\epsilon^{-k} \ket{\tau^a\epsilon^k I}$
    \end{tabular}
  }
  \caption{Unnormalized generalized Dicke states}
      \label{dickesummary}
\end{table}
  \end{center}

\subsection{$G$-symmetrization}

Here is a construction that produces $G$-invariant states.
Let $G$ be a subgroup of $S_n$ and let $t\colon G\to U(1)$ be a
group homomorphism. For 1-qubit states $\ket{\phi_1},\ldots,
\ket{\phi_n}$, we define the $G$-symmetrization of
$\ket{\phi_i}_{i=1}^n$ by
\begin{align}\label{gsymdef}
  &\GSym_t(\ket{\phi_1},\ldots,\ket{\phi_n})\\ \nonumber
  &= K \sum_{\sigma \in G}t_\sigma^{-1}
\ket{\phi_{\sigma^{-1}(1)}}\otimes \cdots \otimes \ket{\phi_{\sigma^{-1}(n)}}
\end{align}
where $K$ is a normalizing factor. (It is possible that
$G$-symmetrization produces the zero vector, which is not a state. For
example, $\GSym_t(\ket{0},\ket{0})=0$ for $G=S_2$ and $t\colon S_2\to
U(1)$ given by $t_{(12)}=-1$. In this case $K$ can be assigned
arbitrarily.) The following Proposition expresses that the
$G$-symmetrization of 1-qubit states is in fact $G$-invariant. The proof
is in the Appendix.

\begin{proposition}\label{gsymisginv}
  Let $\ket{\psi}=\GSym_t(\ket{\phi_1},\ldots,\ket{\phi_n})$
  for some $G$, $t$, and states $\ket{\phi_i}$. Then we have
  $$\sigma \ket{\psi} = t_\sigma\ket{\psi}
  $$
  for all $\sigma \in G$, i.e., $\ket{\psi}$ is $G$-invariant. 
\end{proposition}

\begin{corollary} Suppose that
  $$\ket{\psi} =
  \GSym_t(\ket{\phi_1},\ldots,\ket{\phi_n})\neq 0$$ for some subgroup $G$
  of $S_n$ for some $n\geq 3$, and that $t$ is not trivial. Then
  $\ket{\psi}$ is {\em not} $S_n$-invariant.
\end{corollary}

{\bf Comments.} For $G=S_n$ and for the trivial homomorphism $t$, the
symmetrization~(\ref{gsymdef}) establishes a one-to-one correspondence
between sets of $n$ points on the Bloch sphere and $S_n$-symmetric
states of $n$ qubits.
$$ \{\ket{\phi_1},\ldots,\ket{\phi_n}\} \longleftrightarrow
  \GSym_t(\ket{\phi_1},\ldots,\ket{\phi_n})
$$
This remarkable fact is the well-known Majorana
representation~\cite{bastin2009}. We show with examples that the
one-to-one correspondence does
not hold in general for symmetrization over subgroups of $S_n$.

{\bf Examples:} Unlike the case for $S_n$, $G$-symmetrization does not give
a one-to-one correspondence between $n$-tuples of points on the Bloch
sphere and $G$-invariant states.\\
For $G=A_3=C_3=\langle (123)\rangle$ and $t$ determined by
$t_{(123)}=e^{4\pi i/3}$, and
$$ \ket{\alpha}=\frac{1}{\sqrt{3}}\left(\ket{100}+\omega\ket{010}+\omega^2\ket{001}\right)$$
we have
\begin{align*}
  \ket{\alpha} &= \GSym_t(\ket{1},\ket{0},\ket{0})\\
  &= \GSym_t(\ket{0}+\ket{1},\ket{0}+\omega\ket{1},\ket{0}+\omega^2\ket{1}).
\end{align*}
For $G=A_4$, and $t\colon A_4 \to U(1)$ given by $t_{(123)}=e^{2\pi
  i/3}$, we have
\begin{align*}
  \ket{M_4} &= \GSym_t(\ket{1},\ket{1},\ket{0},\ket{0})\\
  &= \GSym_t(\ket{+},\ket{+},\ket{-},\ket{-})\\
  &= \GSym_t(\ket{1},\ket{1},\ket{+},\ket{+}).
\end{align*}
We explore the structure of $A_n$-symmetrizations
in Subsection~\ref{ansymmsubsect} below.

\section{$A_n$-invariant states}\label{ansection}

\subsection{Phase homomorphisms for $A_n$-invariant states}\label{andickesubsection}

We start by showing that the phase homomorphism $t\colon A_n\to U(1)$
cannot be trivial for a state that is $A_n$-invariant but {\em not} $S_n$-invariant.

\begin{lemma}\label{antrivisn}
% \label{annotsnnottrival}
Let $n\geq 3$, and let $\ket{\psi}$ be
$A_n$-invariant. Suppose $\sigma\ket{\psi}=\ket{\psi}$ for all $\sigma$
in $A_n$. Then $\ket{\psi}$ is $S_n$-invariant.
\end{lemma}

{\bf Proof.} Suppose there is an $I$ such that $1\leq \wt I\leq n-1$ and
$c_I\neq 0$. Choose $j,k,\ell$ such that $i_k=i_\ell$ and $i_j\neq
i_\ell$. The we have $I = (jk\ell)(k\ell)I$, but $(jk\ell)(k\ell)=(kj)$
and $(kj)I\neq I$. Therefore $\wt I=0$ or $\wt I=n$. Thus $\ket{\psi}$
is $S_n$-invariant.

\begin{corollary}\label{annotsnnottrival}
  Suppose that $\ket{\psi}$ is an $A_n$-invariant but not
  $S_n$-invariant state for $n\geq 3$, and suppose that $\sigma \ket{\psi}
  = t_\sigma \ket{\psi}$ for all $\sigma$ in $A_n$. Then there exists a
  3-cycle $\tau$ such that $t_\tau\neq 1$.
\end{corollary}

{\bf Proof.} The Corollary follows directly from Lemma~\ref{antrivisn} with the
observation that $A_n$ is generated by its 3-cycles.

The proofs of the next two statements, Corollary~\ref{tvaluesan} and
Proposition~\ref{noa5inv}, are in the Appendix.

\begin{corollary}\label{tvaluesan}
  {\bf ($t_\sigma$ values for 3-cycles and products of disjoint
    2-cycles)} For a 3-cycle $\sigma$, we have $t_\sigma=1,\omega,
  \omega^2$, where $\omega=e^{2\pi i/3}$. If $\tau$ is a a
  product of disjoint 2-cycles, then $t_\tau=1$.
\end{corollary}

\begin{proposition}\label{noa5inv}
  For $n\geq 5$, there are no pure states that are
  $A_n$-invariant and not $S_n$-invariant.
\end{proposition}

Now we use the Dicke form~(\ref{dickeform}) and the above results about
the phase homomorphism $t\colon A_n\to U(1)$ to determine the form of a state $\ket{\psi}$
that is $A_n$-invariant and not $S_n$-invariant. By Proposition~\ref{noa5inv}, we need only
consider the cases $n=3$ and $n=4$. For $n=3$, the homomorphism $t$ is
determined by choosing one of the two values
$t_{(123)}=\omega,\omega^\ast$, where $\omega=e^{2\pi i/3}$. If the
  coefficient of $\ket{100}$ in the expansion $\ket{\psi}=\sum_I\ket{I}$
  is $c_{100}$, then the weight 1 terms in $\ket{\psi}$ must have the form
  $$c_{100}\left(\ket{100}+\omega\ket{010}+\omega^2\ket{001}\right)$$
  or the conjugate of that expression, depending on the value of
  $t_{(123)}$. A similar observation holds for the weight 2 terms. It is
  easy to see that the weight 0 and weight 3 terms must be zero.
      Let us define the following 3-qubit states, where $\omega =
      e^{2\pi i/3}$ and $a,b$ are some complex
constants with $|a|^2+|b|^2=1$.
      \begin{align*}
        \ket{\alpha}&=\frac{1}{\sqrt{3}}\left(\ket{100}+\omega\ket{010}+\omega^2\ket{001}\right)\\
        \ket{\beta}&=\frac{1}{\sqrt{3}}\left(\ket{110}+\omega\ket{011}+\omega^2\ket{101}\right)\\
     \ket{M_3(a,b)}&=a\ket{\alpha}+b\ket{\beta}   
      \end{align*}
The discussion in the previous paragraph establishes the following.

  \begin{proposition} \label{M3}
If $\ket{\psi}$ is $A_3$-invariant and not
$S_3$-invariant, then there are complex numbers $a,b$ with
$|a|^2+|b|^2=1$ such that $\ket{\psi}$ equals $\ket{M_3(a,b)}$ or its
conjugate, up to a global phase factor.
\end{proposition}

For $n=4$, any permutation $\sigma$ in $A_4$ 
is either the product of disjoint transpositions $\sigma = (ij)(k\ell)$
or a 3-cycle $\sigma=(abc)$. For a 3-cycle, we have $(abc)^3=e$, so we
must have $t_{(abc)}=\omega,\omega^\ast$, where again, $\omega=e^{2\pi
  i/3}$. In the first case, we have
$$\sigma=(ij)(k\ell) =(i\ell k)(ijk)
$$
Thus, for $\sigma=(ij)(k\ell)=(i\ell k)(ijk)$, we must have $t_\sigma$
is a power of $\omega$. But $\sigma^2=e$ implies $t_\sigma = \pm
1$. Therefore we must have $t_\sigma=1$, since no power of $\omega$ can
equal $-1$. The values of $t_\sigma$ for $\sigma=(abc)$ are determined
by equations like $(123)(124)=(13)(24)$, so $t_{(123)}=
t_{(124)}^\ast$. With just a few calculations, we see that the values
$t_\sigma$ must either be the same as those given in
Table~\ref{tab:a4m4actvalues}, or their conjugates. If $\ket{\psi}$ is
$A_4$-invariant but not $S_4$-invariant, it cannot have any weight 1
terms with nonzero coefficients in its expansion in the computational
basis. For example, the 2-cycle $(123)$ fixes $\ket{0001}$, but
$t_{(123)}\neq 1$. By similar considerations, $\ket{\psi}$ cannot have
  any nonzero terms in weights 0, 1, 3, or 4. One determines quickly
  that the weight 2 terms must be organized as for $\ket{M_4}$ or its
  conjugate (up to a global phase factor). 
Thus we have proved
the following.

\begin{proposition}\label{M4prop}
  If $\ket{\psi}$ is $A_4$-invariant and not
  $S_4$-invariant, then $\ket{\psi}$ equals $\ket{M_4}$ or its conjugate, up to
  a phase multiple.
\end{proposition}

\subsection{Local unitary equivalence}

In this subsection we classify the LU equivalence classes of states that
are $A_n$-invariant and not $S_n$-invariant. For the case $n=3$, we have
the striking result that all of the states in the infinite family
$\ket{M_3(a,b)}, \ket{\overline{M_3(a,b)}}$ are local unitary
equivalent. The proof exploits the relationship between carefully chosen
1-qubit operators $U$ acting on the 1-qubit state
$a\ket{0}+b\ket{1}$ and suitably constructed 3-qubit operators $V^{\otimes 3}$ acting on
$\ket{M_3(a,b)}$, in such a way that a local equivalence between
$a\ket{0}+b\ket{1}$  and $a'\ket{0}+b'\ket{1}$ yields a local
equivalence between $\ket{M_3(a,b)}$ and $\ket{M_3(a',b')}$.
A detailed proof is in the Appendix.

\begin{theorem}\label{a3oneluclassthm}
  Let $\ket{\psi}$ be $A_3$-invariant and not $S_3$-invariant. Then
  $\ket{\psi}$ is LU equivalent to
  $$\ket{M_3(1,0)} =
  \frac{1}{\sqrt{3}}\left(\ket{100}+\omega\ket{010}+\omega^2\ket{001}\right).
  $$
\end{theorem}

For the case $n=4$, there are only two states to consider, namely
$\ket{M_4}$ and its conjugate. We refer the reader to~\cite{maxstabnonprod2}
for a proof that there are LU invariants that distinguish the LU classes
of $\ket{M_4}$ and its conjugate. We record the result here.

\begin{proposition}\label{a4twoluclasses}
The two states (up to global phase factor) that are $A_4$-invariant and
not $S_4$-invariant, namely, $\ket{M_4}$ and its conjugate, are
LU-inequivalent.
\end{proposition}

\subsection{$A_n$-symmetrization}\label{ansymmsubsect}

In this subsection, we consider the question: under what conditions does
$A_n$-symmetrization produce an $A_n$-invariant state that is not also
$S_n$-invariant and also nonzero? In the Propositions below, we
characterize the $n$-tuples $\ket{\phi_1},\ldots,\ket{\phi_n}$ of
1-qubit states and the homomorphisms $t\colon A_n\to U(1)$ such that,
for $G=A_n$, the state
$\GSym_t(\ket{\phi_1},\ldots,\ket{\phi_n})$ is $A_n$-invariant and not
$S_n$-invariant and also nonzero. The proofs are in the Appendix.

By the results in the preceding
subsection, we need only consider $n=3,4$. 

\begin{proposition}\label{A3symprop}
  Let $n=3$, let $G=A_3$, and let $\ket{\phi_i}$ be 1-qubit states for
  $i=1,2,3$. The state
  $\GSym_t(\ket{\phi_1},\ket{\phi_2},\ket{\phi_3})$ is $A_3$-invariant
  and not $S_3$-invariant and not zero if an only if $t_{(123)}\neq 1$ and the three
  states $\ket{\phi_1},\ket{\phi_2},\ket{\phi_3}$ are not all the same
  (up to phase).
\end{proposition}

\begin{proposition}\label{A4symprop}
  Let $n=4$, let $G=A_4$, and let $\ket{\phi_i}$ be 1-qubit states for
  $i=1,2,3,4$. Except for a set of measure zero, the state
  $\GSym_t(\ket{\phi_1},\ket{\phi_2},\ket{\phi_3},\ket{\phi_4})$ is $A_4$-invariant
  and not $S_4$-invariant and not zero if an only if $t_{(123)}\neq 1$
  and no three of the states
  $\ket{\phi_1},\ket{\phi_2},\ket{\phi_3},\ket{\phi_4}$ are equal (up to
  phase).
\end{proposition}

\section{$C_n$ and $D_n$-invariant states}\label{cndnsection}

\subsection{Necklace diagrams}

\begin{figure}
  \centering{
  \includegraphics[scale=.30]{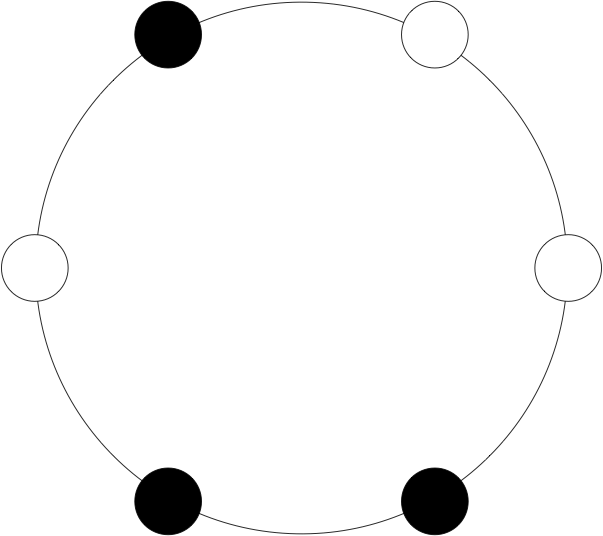}
  \caption{Necklace diagram for the cycle class $[101100]$}
  \label{necklace_eg}
  }
\end{figure}

A regular $n$-gon with vertices colored white or black encodes the cycle
class of an $n$-bit string, as follows. Starting at any vertex, label
the vertices $v_1,v_2,\ldots, v_n$ traveling counterclockwise around the
polygon. Let the bit string $I=i_1i_2\ldots i_n$ be defined by $i_k=0$
if $v_k$ is white, and let $i_k=1$ if $v_k$ is black. If we perform the
same procedure starting at a different vertex, say $w_1=v_\ell$, we
obtain the bit string $J= \epsilon^\ell I$, where $\epsilon = (12\cdots
n)$. This type of figure, called a {\em necklace diagram}, encodes the
$C_n$ orbit $[I]$ of the bit string $I$. Figure~\ref{necklace_eg}
illustrates with an example.

\begin{figure}
  \centering{
  \includegraphics[scale=.65]{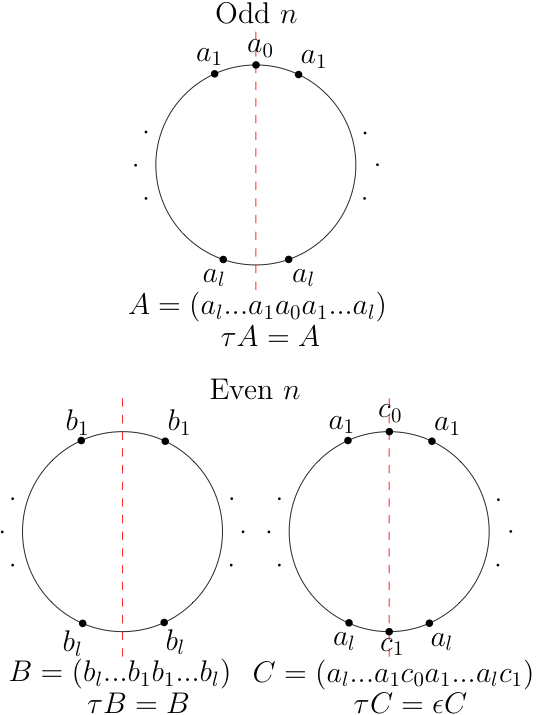}
  \caption{Lines of symmetry through 0, 1, and 2 vertices}
  \label{Necklace_Diagrams}
  }
\end{figure}

In what follows, we will develop criteria for cyclic and dihedral
symmetry for states constructed from necklace diagrams in terms of lines
of mirror symmetry. For $n$ odd, any line of mirror symmetry passes through one
vertex, say $v_0$, labeled by a bit $a_0$, and one of the bit strings in the cycle class of the
necklace has the form $A=(a_\ell \ldots a_2a_1a_0a_1a_2\ldots a_{\ell})$
so that $\tau A=A$. For $n$ even, there are two possible types of mirror
lines of symmetry: a line may pass through no vertices, and a line may
pass through two vertices. In the no vertex case, the necklace encodes
a palindromic bit string of the form $B=(b_1b_2\ldots b_\ell b_\ell
\ldots b_2b_1)$ so that $\tau B=B$. In the two vertex case, the necklace encodes a bit
string of the form $C=(a_\ell \cdots a_2a_1c_0 a_1a_2\ldots a_\ell
c_1)$, so that $\tau C=\epsilon C$. See Figure~\ref{Necklace_Diagrams}.

Terminology: We will use the following terms for types of necklaces. A
necklace $[I]$ of length $n$ is:
\begin{itemize}
\item [{[\SP]}] {\em self-palindromic} or {\em palindromic on the string
  level}, denoted \SP, if $[I]$ has at least one 0-vertex line of
  symmetry (if $n$ is even) or at least one 1-vertex line of symmetry
  (if $n$ is odd)
  \item [[{\CP]}] {\em class-palindromic} or {\em palindromic on the
    class level}, denoted \CP, if $n$ is even and $[I]$ has at least one
    2-vertex line of symmetry and no 0-vertex lines of symmetry
    \item [{[chiral]}] {\em chiral} or {\em non-palindromic}, if $[I]$
      has no lines of symmetry
\end{itemize}
We say that the {\em cycle order} of a necklace diagram $[I]$ is the
number of distinct bit strings the $G$-orbit $[I]$. Equivalently, we can
define the cycle order of a bit string $I$ to be the smallest positive
integer $m$ such that $\epsilon^m I=I$.

The following Proposition will be used in the characterization of states
that are $C_n$-invariant and not $D_n$-invariant. The proof is in the Appendix.
\begin{proposition}\label{spevencpodd}
  Let $J$ be an $n$-bit string with even cycle order, say m.
  Suppose $\tau\epsilon^k J = J$. If $[J]$ is of type \SP\, then $k$ is
  even. If $[J]$ is of type \CP\, then $k$ is odd.
\end{proposition}

We conclude with a geometric observation that relates the cycle order
to the number of lines of mirror symmetry of a necklace diagram. A proof
sketch is in the Appendix.
\begin{proposition}\label{cycleorderbymirrorlinect}
  Let $L$ be the number of lines of mirror symmetry of an $n$-bit necklace
  diagram $[I]$. If $L=0$, then the cycle order of $[I]$ is $n$. If
  $L>0$, then the cycle order of $[I]$ is $n/L$.
\end{proposition}

\subsection{$C_n$ and not $D_n$-invariant states}

In this subsection, we characterize those states that are
$C_n$-invariant and are {\em not} $D_n$-invariant. The proof of
Theorem~\ref{cnnotdninv} is in the Appendix.

\begin{proposition}\label{cntextendtodn}
  Let $\ket{\psi}$ be $C_n$-invariant so that we have
  $$\epsilon^k\ket{\psi} = t_\epsilon^k \ket{\psi}
  $$
  for all $k$. The state $\ket{\psi}$ is $D_n$-invariant if and only if
  $$ \tau \ket{\psi}=t_\tau \ket{\psi}
  $$
  for some $t_\tau$.
\end{proposition}

{\bf Proof.} This follows from the fact that all elements of $D_n$ can
be written in the form $\tau^a \epsilon^k$ for some $a=0,1$ and some
integer $k$.

\begin{theorem}\label{cnnotdninv}
  Let $\ket{\psi}$ be a $C_n$-invariant state with
  homomorphism $t\colon C_n\to U(1)$ determined by 
  $\epsilon\ket{\psi}=t_\epsilon \ket{\psi}$. The state $\ket{\psi}$ is
  $D_n$-invariant, with homomorphism $s\colon D_n\to U(1)$ determined by
  $\sigma\ket{\psi}=s_\sigma \ket{\psi}$ for all $\sigma\in D_n$
  if and only
  if all four of the following hold:
  \begin{enumerate}
  \item [(i)] $t_\epsilon = \pm 1$
  \item [(ii)] if there is an $I$ of type \SP\ with $c_{[I]}\neq 0$, then $s_\tau=1$
  \item [(iii)] if there is an $I$ of type \CP\ with $c_{[I]}\neq 0$, then $s_\tau=s_\epsilon$
  \item [(iv)] if there is an $I$ of chiral type with $c_{[I]}\neq 0$,
    then $c_{J}=s_\tau c_{\tau J}$ for all $J\in [I]$
  \end{enumerate}
\end{theorem}

\subsection{$D_n$ and not $S_n$-invariant states}

The following Proposition characterizes those states that are
$D_n$-invariant and are {\em not} $S_n$-invariant. The proof is in the Appendix.

\begin{proposition}\label{dnnotsninv}
  Suppose $\ket{\psi}$ is $D_n$-invariant. Then $\ket{\psi}$ is
  $S_n$-invariant if and only if both of the following hold.
  \begin{enumerate}
  \item [(i)] $t_\epsilon=t_\tau=1$
  \item [(ii)] $C_{[I]}=C_{[J]}$ for all $I,J$ with $\wt I= \wt J$.
  \end{enumerate}
\end{proposition}

\subsection{Local unitary equivalence}

The full characterization of local unitary equivalence classes for
$C_n$-invariant and $D_n$-invariant states (that are not
$S_n$-invariant) awaits future work. The fact that there is only one
local unitary class for the case of $A_3=C_3$
(Theorem~\ref{a3oneluclassthm}) suggests that the general case will be
subtle. We record here a preliminary result for local unitary classes of
the generalized Dicke states for $C_n$-invariant and $D_n$-invariant
states, {\em whether or not} they are also $S_n$-invariant. 

We show in~\cite{symmstatespaper} that the Dicke states $\ket{D_n^w}$
(eqn.~(\ref{sndicke})) for
$S_n$-invariant states belong to $\lfloor n/2\rfloor$ distinct local
unitary classes, one for each weight $w=0,1,2,\ldots,\lfloor n/2\rfloor$
The Dicke state $\ket{D_n^w}$ is local unitary
equivalent to the Dicke state $\ket{D_n^{n-w}}$ via the
operator $X^{\otimes n}$. The same proof, mutatis mutandis, can be used
for the subgroups $C_n$ and $D_n$. We give a sketch of the proof in the
Appendix.

\begin{proposition}
  \label{weightsludistinct}
  Let $\ket{D_t^{[I]}}, \ket{D_t^{[J]}}$ be generalized Dicke states
  (see Table~\ref{dickesummary}) for
  $G=C_n$ or $G=D_n$. If $\wt J\neq \wt I$ and $\wt J\neq n-\wt I$, then
  the states $\ket{D_t^{[I]}}$ to $\ket{D_t^{[J]}}$ belong to distinct
  local unitary classes.
\end{proposition}

\section{Applications}\label{applicationssection}

{This section provides evidence, both established and conjectural, that
it is reasonable to expect to find resource states for quantum
information protocols among the $G$-invariant states, for $G\subset
S_n$.}

\subsection{Quantum codes}
{In the Introduction, we describe a 4-qubit code that
  uses the logical qubit states
  \begin{align*}
    \ket{0_L} &= \ket{M_4}\\
    \ket{1_L} &= \ket{\overline{M}_4}
  \end{align*}
where $\ket{\overline{M}_4}=(12)\ket{M_4}$. Owing to the local unitary
invariance of these states, {that is,
  \begin{align*}
    U^{\otimes 4}\ket{M_4} &= \ket{M_4}\\
    U^{\otimes 4}\ket{\overline{M}_4} &= \ket{\overline{M}_4}
  \end{align*}
  for all 1-qubit unitary operators $U$~\cite{maxstabnonprod2},
this code is unaffected by noise evolutions 
of the form $U^{\otimes 4}$}. Here is
a construction in a $G$-invariant state framework that generalizes this
example. Let $\ket{\psi}$ be a $G$-invariant state for some group of
permutations $G$. The $S_n$-orbit $\{\sigma\ket{\psi}\colon \sigma\in
S_n\}$ of $\ket{\psi}$ may contain a set of orthogonal states. (For
example, for the 6-qubit $C_n$-invariant state
$$\ket{\psi}=\sum_{k=0}^5 e^{2\pi i k/6}\epsilon^k\ket{000001}$$
where $\epsilon$ is the 6-cycle $\epsilon=(123456)$, the state
$(12)(34)(56)\ket{\psi}$ is orthogonal to $\ket{\psi}$, but
$(12)\ket{\psi}$ is not.) We leave it to future work to study properties
of codes that use orthogonal sets of codewords in $S_n$ orbits of
$G$-invariant states.}

\subsection{Hypergraph states with GHZ-like local unitary stabilizers}

{The GHZ state of $n$-qubits
$$\frac{1}{\sqrt{2}}\left(\ket{0}^{\otimes n} + \ket{1}^{\otimes n}\right)
$$ is an important resource in quantum information
  theory. In~\cite{hypergraph2014} (see Prop.~5.5), we construct
  hypergraph state $\ket{\psi}$ consisting of $n$ ``essential'' qubits
  together with an auxiliary core of an arbitrary number $m$ of qubits
  in such a way that $\ket{\psi}$ is invariant under permutations of the
  auxiliary qubits. Briefly, $\ket{\psi}$ is constructed by starting
  with the uniform superposition $H^{\otimes (m+n)}\ket{0}^{\otimes
    (m+n)}$ of all the computational basis states and then applying, for
  each essential qubit $k$, the operator
$$
  C_k=  \1 - 2(\ket{1}\bra{1})^{\otimes (m+1)}
$$
to the $m$ auxiliary qubits together with the
$k$-th essential qubit:
$$\ket{\psi} = \left(\prod_{k=1}^n C_k\right) H^{\otimes (m+n)}\ket{0}^{\otimes (m+n)}.
$$ The state $\ket{\psi}$ shares entanglement properties with the
$n$-qubit GHZ state in the sense that the local unitary stabilizer group
of $\ket{\psi}$ is isomorphic to the local unitary stabilizer group of
the GHZ state, where the essential qubits of $\ket{\psi}$ correspond to
the qubits of the GHZ state. Results await future investigation, but we
expect that the ``GHZ-like'' state $\ket{\psi}$ will serve as a useful
version of a GHZ state, for example, in a setting where there is a
possibility of the loss of some qubits. We conjecture that protocols for
the GHZ state that are based on the local unitary stabilizer group will
adapt to $\ket{\psi}$. An example is a verification protocol (Pappa et
al.~\cite{pappa2012}), based on applying random local unitary
stabilizers, that verifies (or disqualifies) possibly untrusted GHZ
states whose qubits are distributed among parties who may or may not be
trusted.}

\subsection{Further applications}

{Burchardt et al.~\cite{burchardt2021} describe
how $G$-invariant states (called ``Dicke-like'' states in their
paper), can be used for parallel teleportation protocols, secret sharing
schemes, and quantum chemistry applications to molecules with a high degree
of spatial symmetry. The authors demonstrate how entanglement in
Dicke-like states, as measured by concurrence, can be concentrated
between selected subsets of parties, while at the same time suppressing
correlations between other pairs of parties. The authors conjecture that
this property will be useful for protocols where variable strength of
entanglement interactions between certain parties is desired.}

\section{Conclusion}

{We have characterized multiqubit states that are invariant under the
alternating, cyclic, and dihedral subgroups of the group of permutations
of the qubits, and have described applications to quantum technology
that exploit these symmetries. }

Directions for continued work on the
classification of states that are invariant under
subgroups of $S_n$ include the following. 
Can we characterize the configurations of Bloch sphere points that have the
same $G$-symmetrizations? Could we use such a characterization to prove
things about local unitary classes of $G$-invariant states (like we can for $S_n$-invariant
states)?
%% What about other subgroups $G$ of $S_n$? For example, certain
%% hypergraph state constructions are invariant under $G=S_m \times
%% S_{m'}$, and may have applications for quantum computational
%% protocols~\cite{hypergraph2014}.

{Towards applications, what performance properties are possessed by codes
that are invariant under proper subgroups of $S_n$? How can the loss
tolerance of the ``GHZ-like'' state (described in the previous section)
enhance entanglement verification protocols?}

Finally, as we have done
for symmetric states~\cite{symmstates2}, it will be natural to extend
this work to mixed states.

\subsection*{Acknowledgment}
This work was supported by National Science Foundation grant
PHY-2011074. We are grateful for stimulating conversations with Adam Burchardt.

%\addcontentsline{toc}{section}{References}    
%\section{References}
%\bibliographystyle{unsrt}
%\bibliography{groupnotes}
%\bibliography{dwl_main}

    \appendix

    \section{Proofs of propositions}

{\bf Proof of Proposition~\ref{singletonlyphaseinv}.} By the properties of group actions, we have $t_e=1$, where
$e$ is the identity permutation, and $t_{\sigma\tau} = t_\sigma t_\tau$
for all $\sigma,\tau$ in $S_n$. In particular, we may write any
permutation $\sigma$ as a product 
$$\sigma = \tau_1\tau_2\cdots \tau_k$$
of transpositions $\tau_j$. Thus it suffices to show that $t_\tau=1$ for
any transposition $\tau$ in order to prove the proposition. The case
$n=1$ is trivial, so now assume that $n\geq 3$. Let $\tau=(k\ell)$ be a
transposition. Write $\ket{\psi} = \sum_I c_I\ket{I}$ in the standard
basis, and choose a multi-index $I_0=i_1i_2\cdots i_n$ such that
$i_k=i_\ell$ and $c_{I_0}\neq 0$. (Start by choosing any $J$ such that
$c_J\neq 0$. By the pigeonhole principle, there must be two indices
$a,b$ such that $j_a=j_b$. Apply a permutation $\sigma$ that takes $a,b$
to $k,\ell$, and let $I_0=\sigma J$.) Now we have
$$\sigma \ket{\psi} = c_{I_0}\ket{I_0} + \sum_{J\neq I_0}
t_\tau c_J\ket{\tau J}.
$$
But this last expression must equal $\ket{\psi}= c_{I_0}\ket{I_0} +
\sum_{J\neq I_0} c_J\ket{J}$, so it must be that $t_\tau = 1$.

{\bf Proof of Proposition~\ref{dickeformisginv}.} First, observe that
the requirement that $t$ is constant on stabilizers of strings $I$ with
nonzero coefficients $c_I$ guarantees that the expression $t_g$ does not
depend on which $g\in G$ is chosen in the expression $t_g$, as long as
$gL_{[I]}=J$. Indeed, suppose that $gL = hL$. Then $g^{-1}hL=L$, so
$t_g^{-1}t_h = t_e=1$, so $t_g=t_h$.

Now, let $h\in G$. We have
\begin{align*}
  h\ket{\psi} &= \sum_{[I]} c_{[I]} \sum_{J\in [I]} t_g^{-1}\ket{J}\\
  &= h\left(\sum_{[I]} c_{[I]} \sum_{J\in [I]} t_{(h^{-1}g)}^{-1}\ket{h^{-1}J}\right)\\
  &= \sum_{[I]} c_{[I]} \sum_{J\in [I]} t_{(h^{-1}g)}^{-1}\ket{J}\\
  &= \sum_{[I]} c_{[I]} \sum_{J\in [I]} t_{h}t_{g}^{-1}\ket{J}\\
  &= t_h\ket{\psi}.
\end{align*}

{\bf Proof of Proposition~\ref{gsymisginv}.} We have
\begin{align*}
  \sigma \ket{\psi} &= \sigma \sum_{\pi\in G} t_\pi^{-1}
  \ket{\phi_{\pi^{-1}(1)}} \cdots \ket{\phi_{\pi^{-1}(n)}}\\
&= \sum_{\pi\in G} t_\pi^{-1}
  \ket{\phi_{\pi^{-1}\sigma^{-1}(1)}} \cdots
  \ket{\phi_{\pi^{-1}\sigma^{-1}(n)}}\\
&= \sum_{\pi\in G} t_\pi^{-1}
  \ket{\phi_{(\sigma\pi)^{-1}(1)}} \cdots
  \ket{\phi_{(\sigma\pi)^{-1}(n)}}\\
&= \sum_{\xi\in G} t_{\sigma^{-1}\xi}^{-1} 
  \ket{\phi_{\xi^{-1}(1)}} \cdots
  \ket{\phi_{\xi^{-1}(n)}} \hspace*{.2in}  \mbox{($\xi = \sigma\pi$)}\\
&= t_\sigma \sum_{\xi\in G} t_{\xi}^{-1}
  \ket{\phi_{\xi^{-1}(1)}} \cdots
  \ket{\phi_{\xi^{-1}(n)}}\\
  &= t_\sigma \ket{\psi}.
\end{align*}

{\bf Proof of Corollary~\ref{tvaluesan}.} The first statement follows from $\sigma^3=1$. From
$\tau^2=1$, we have $t_\tau=\pm 1$ for $\tau=(ab)(cd)$ with $a,b,c,d$ distinct. From the equation
$$(ab)(cd) = (dca)(abc)
$$
we have $t_\tau = \omega^k$ for some $k$. It follows that $t_\tau=1$.

Now let $n\geq 3$ and suppose that $\ket{\psi}$ is $A_n$-invariant but not
$S_3$-invariant. By Corollary~\ref{annotsnnottrival}, there is a 3-cycle $(abc)$ such that
$t_{(abc)}\neq 1$. Let $u,v,w$ be 3 distinct values in
$\{1,2,\ldots,n\}$. We have the following equations in $S_n$.
\begin{align}\label{evenoverlap}
  (vuw) &= (uv)(cw)(bv)(au)(abc)(au)(bv)(cw)(uv)\\
  \label{oddoverlap}
  (uvw) &= (cw)(bv)(au)(abc)(au)(bv)(cw)
\end{align}
Consider the three equations
\begin{equation}\label{3possibleoverlaps}
  a=u,b=v,c=w.
\end{equation}
We consider four cases.
\begin{itemize}
\item If none of the equations~(\ref{3possibleoverlaps}) hold,
  then~(\ref{evenoverlap}) is an equation in $A_n$. By
  Corollary~\ref{annotsnnottrival}, it follows that $t_{(uvw)} =
  t_{(abc)}^\ast$.
\item If exactly 1 of equations~(\ref{3possibleoverlaps}) holds,
  then~(\ref{oddoverlap}) is an equation in $A_n$. By
  Corollary~\ref{annotsnnottrival}, it follows that $t_{(uvw)} =
  t_{(abc)}$.
\item If exactly 2 of equations~(\ref{3possibleoverlaps}) hold,
  then~(\ref{evenoverlap}) is an equation in $A_n$. By
  Corollary~\ref{annotsnnottrival}, it follows that $t_{(uvw)} =
  t_{(abc)}^\ast$.
  \item If all 3 of equations~(\ref{3possibleoverlaps}) hold,
  then we have $(uvw)=(abc)$, and so $t_{(uvw)} = t_{(abc)}$.
\end{itemize}

{\bf Proof of Proposition~\ref{noa5inv}.} Let $n\geq 5$ and suppose that $\ket{\psi}$ is
$A_n$-invariant and not $S_n$-invariant. Let
$\ket{\psi}=\sum_Ic_I\ket{I}$ be the expansion of $\ket{\psi}$ in the
computational basis. Choose $I$ such that $c_I\neq 0$. Because $n\geq
5$, there must be three positions $u,v,w$ such that $i_u=i_v=i_w$. By
Corollary~\ref{annotsnnottrival}, there must be some 3-cycle $(abc)$
such that $t_{(abc)}\neq 1$.  By the discussion immediately preceding
the statement of the Proposition, we have $t_{(uvw)}=t_{(abc)}$ or
$t_{(uvw)}=t_{(abc)}^\ast$. In both cases, $t_{(uvw)}\neq 1$. Now
$(uvw)I = I$, but $t_{(uvw)}I \neq I$. Thus we must have $c_I=0$,
contradicting our assumption. It follows that the state $\ket{\psi}$
cannot exist.

  {\bf Another Proof of Proposition~\ref{noa5inv}.} Observe that the first two cases in bulleted
  list in the proof of Corollary~\ref{tvaluesan} both lead to
  contradictions. In place of~(\ref{evenoverlap})
  and~(\ref{oddoverlap}), consider equations
  \begin{align}\label{evenoverlap2}
  (uvw) &= (uv)(cw)(bu)(av)(abc)(av)(bu)(cw)(uv)\\
  \label{oddoverlap2}
  (vuw) &= (cw)(bu)(av)(abc)(av)(bu)(cw)
  \end{align}
  that interchange the 3-cycles $(uvw),(uwv)$ on the left.
In the first case (none of the equations
  in~(\ref{3possibleoverlaps}) hold), apply~(\ref{evenoverlap2}) to
  conclude that $t_{(uvw)}=t_{(abc)}$. But this contradicts the
    conclusion in the proof to Corollary~\ref{tvaluesan} that
    $t_{(uvw)}=t_{(abc)}^\ast$ (because $t_{(abc)}=\omega,\omega^\ast$
      is not real). Similarly, in the second case (exactly one of the equations
  in~(\ref{3possibleoverlaps}) holds), apply~(\ref{oddoverlap2}) to
  conclude that $t_{(uvw)}=t_{(abc)}^\ast$. This is again a
    contradiction to the conclusion in Corollary\ref{tvaluesan} that we have
    $t_{(uvw)}=t_{(abc)}$. Now we have ruled out the possibility of the
      first two cases, so it must be that exactly 2 or 3 of
      equations~(\ref{3possibleoverlaps}) hold. This in turn implies
      that $n=3$ or $n=4$.

\subsection*{Proof of Theorem~\ref{a3oneluclassthm}}      

We begin with a Lemma about rotations of the unit sphere. We will write
$(a,b,c)=(\theta,\phi)_{\mbox{\tiny \rm spherical}}$ to denote the point
on the unit sphere with rectangular coordinates $(a,b,c)$ and spherical
coordinates $(\theta,\phi)$, that is, $a=\cos\phi \sin\theta$,
$b=\sin\phi\sin\theta$, $c=\cos\theta$.  We will write
$R_{\theta,(a,b,c)}$ to denote the rotation of the sphere by $\theta$
radians about the axis determined by the point $(a,b,c)$ on sphere. We
will also write $R_{\theta,Z},R_{\theta,Y}$ to denote the rotations
$R_{\theta,(0,0,1)},R_{\theta,(0,1,0)}$, respectively.

\begin{figure}
  \centering{
  \includegraphics[scale=.6]{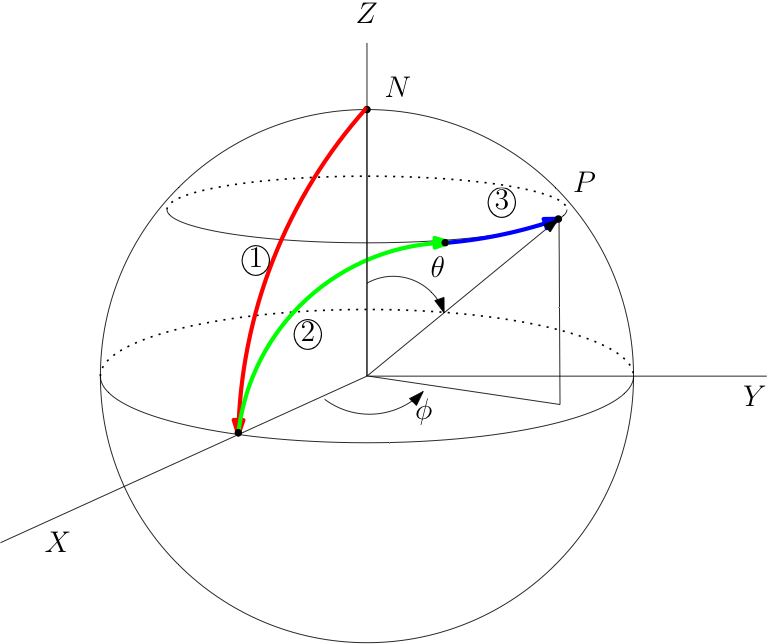}
  \caption{Moving $N$ to $P$ with rotations
    $\mbox{\textcolor{red}{\textcircled{1}}}=  R_{\frac{\pi}{2},Y}$,
    $\mbox{\textcolor{green}{\textcircled{2}}}=  R_{-\frac{\pi}{2},(\cos(\frac{\pi}{2}-\theta),\sin(\frac{\pi}{2}-\theta),0)}$,
    $\mbox{\textcolor{blue}{\textcircled{3}}}=  R_{\phi-(\frac{\pi}{2}-\theta),Z}$
  }
  \label{3rotsfig}
  }
\end{figure}

\begin{lemma}\label{3rotslemma}
  Let $(\theta,\phi)$ be the spherical coordinates of a point
  $P$ on the sphere, so that
  $$P=(\sin\theta\cos\phi,\sin\theta\sin\phi,\cos\theta).$$ Let
  $N=(0,0,1)$ and let
  \begin{equation}\label{NtoProtation}
    R = R_{\phi-(\frac{\pi}{2}-\theta),Z} \circ
    R_{-\frac{\pi}{2},(\cos(\frac{\pi}{2}-\theta),\sin(\frac{\pi}{2}-\theta),0)} \circ R_{\frac{\pi}{2},Y}.
  \end{equation}
  The we have
  $$R(N)=P.
  $$
  See Figure~\ref{3rotsfig}.
\end{lemma}

{\bf Proof.} We have
\begin{align*}
  &R(N)\\
  &= \left(R_{\phi-(\frac{\pi}{2}-\theta),Z} \circ R_{-\frac{\pi}{2},(\cos(\frac{\pi}{2}-\theta),\sin(\frac{\pi}{2}-\theta),0)} \circ
  R_{\frac{\pi}{2},Y}\right)(N)\\
  &=
  \left(R_{\phi-(\frac{\pi}{2}-\theta),Z} \circ R_{-\frac{\pi}{2},(\cos(\frac{\pi}{2}-\theta),\sin(\frac{\pi}{2}-\theta),0)}\right)(1,0,0)\\
  &=
  R_{\phi-(\frac{\pi}{2}-\theta),Z} \left(\theta,\frac{\pi}{2}-\theta\right)_{\mbox{\tiny \rm
      spherical}}\\
  &=(\theta,\phi)_{\mbox{\tiny \rm
      spherical}}\\
  &=P.
\end{align*}
This proves the Lemma.

\begin{lemma}\label{exppiby4xyaxis}
  Let $U=e^{i\pi/4(\alpha X+ \beta Y)}$, where $\alpha,\beta$
  are real and $\alpha^2+\beta^2=1$, and let $\xi =
  \alpha+i\beta$.  We claim that
  $$ U^{\otimes 3} \ket{M_3(a,b)}= \ket{M_3(a',b')}
  $$
  where $(a',b')$ is given by
  \begin{equation} \label{fancymatrix}
    \begin{bmatrix}
      a'\\b'
    \end{bmatrix}
    = \frac{1}{\sqrt{2}}
    \begin{bmatrix}
      1 & e^{i(\pi/2 - \xi - \pi/3)}\\
      e^{i(\pi/2 + \xi + \pi/3)} & 1
    \end{bmatrix}\begin{bmatrix}
      a\\b
    \end{bmatrix}.
  \end{equation}
\end{lemma}

{\bf Proof of Lemma~\ref{exppiby4xyaxis}.} We have
$$U=\exp(i\pi/4(\alpha X+ \beta Y) = \frac{1}{\sqrt{2}}
\begin{bmatrix}
  1 & -e^{-i\phi}\\
  e^{i\phi} & 1
\end{bmatrix}
$$ where $e^{i\xi}=\alpha+i\beta$, and where $\phi=\xi + \pi/2$. Thus
$U$ is the rotation $R_{-\pi/2,(\alpha,\beta,0)}$ of the Bloch sphere
(see~\cite{nielsenchuang},~Exercise 4.6).
Using $$\ket{M_3(a,b)}=\frac{1}{\sqrt{3}} \left(0, a\omega^2, a\omega,
b\omega, a, b\omega^2, b, 0\right)^T,$$ it is straightforward to check
that
$$U^{\otimes 3}\ket{M_3(a,b)} = \frac{1}{\sqrt{2}}
\begin{bmatrix}
  1 & \omega e^{-i\phi}\\
  - \omega^2 e^{i\phi} & 1
\end{bmatrix}
\begin{bmatrix}
  a\\
  b\\
\end{bmatrix}.
$$
Using $\omega=e^{2\pi i/3}$ and $\phi = \xi+\pi/2$, we obtain equation~(\ref{fancymatrix}).

\begin{lemma}\label{z3onm3}
  Let $U=e^{-i\frac{u}{2}Z}$, where $u$ is a real number. We have
  $$U^{\otimes 3} \ket{M_3(a,b)} = \ket{M_3(a',b')}
  $$  where $(a',b')$ is given by
  \begin{equation} \label{fancymatrix2}
    \begin{bmatrix}
      a'\\b'
    \end{bmatrix}
    =  \begin{bmatrix}
      e^{-iu/2} & 0\\
      0  & e^{iu/2}
    \end{bmatrix}\begin{bmatrix}
      a\\b
    \end{bmatrix}.
  \end{equation}
\end{lemma}

{\bf Proof of Theorem~\ref{a3oneluclassthm}.} We will exhibit an LU transformation that takes
$\ket{M_3(1,0)}$ to $\ket{M_3(a,b)}$ for any $(a,b)$. Let $\theta,\phi$
be spherical coordinates for the point point on the Bloch sphere that represents the
1-qubit state $a\ket{0}+b\ket{1}$, that is, we have
$$a\ket{0}+b\ket{1} = \cos\frac{\theta}{2}\ket{0} + e^{i\phi}\sin\frac{\theta}{2}\ket{1}.
$$
Let $U=U_3U_2U_1$, where $U_1,U_2,U_3$ are given by 
  \begin{align}
    U_3 &= \exp\left[\frac{-i}{2}(\phi+\theta-\frac{\pi}{2})Z\right]\\
    U_2 &=   \exp\left[i\frac{\pi}{4}(\cos\left(\frac{\pi}{6}-\theta\right)X +
      \sin\left(\frac{\pi}{6}-\theta\right)Y)\right]\\
    U_1 &=  \exp\left[-i\frac{\pi}{4}(\cos \left(\frac{\pi}{6}\right) X + \sin\left(\frac{\pi}{6}\right)Y)\right].
  \end{align}
Applying Lemmas~\ref{exppiby4xyaxis} and~\ref{z3onm3}, we have
  that $U$ acts as the rotation~(\ref{NtoProtation})
$$R= R_{\phi-(\frac{\pi}{2}-\theta),Z} \circ R_{-\frac{\pi}{2},(\cos(\frac{\pi}{2}-\theta),\sin(\frac{\pi}{2}-\theta),0)} \circ
  R_{\frac{\pi}{2},Y}.
  $$
Thus, by Lemma~\ref{3rotslemma}, we have 
$$U^{\otimes 3} \ket{M_3(1,0)} = \ket{M_3(a,b)},$$
as desired. Now if we are given $\ket{M_3(a',b')}$, use the same
construction to choose a unitary $V$ such that 
$$V^{\otimes 3} \ket{M_3(1,0)} = \ket{M_3(a',b')}.$$
Now we have the LU equivalence
$$(VU^\dagger)^{\otimes 3}\ket{M_3(a,b)}= \ket{M_3(a',b')}.$$

Finally, we show that $\overline{\ket{M_3(a,b)}}$ is LU equivalent
to $\ket{M_3(1,0)}$. Given $a,b$, construct an LU operator $U$ as above
such that
$U\ket{M_3(1,0)}=\ket{M_3(\overline{a},\overline{b})}$. Then apply the
LU operator
$$
W=
\begin{bmatrix}
  1 & 0\\
  0 & 1
\end{bmatrix}
\otimes
\begin{bmatrix}
  1 & 0\\
  0 & \omega
\end{bmatrix}
\otimes
\begin{bmatrix}
  1 & 0\\
  0 & \omega^2
\end{bmatrix}
$$
to obtain the LU equivalence
$$ WU\ket{M_3(1,0)}=\overline{\ket{M_3(a,b)}}.
$$

{\bf Proof of Proposition~\ref{A3symprop}.}

Let $G=A_3$, and let $\ket{\psi}=\GSym_t(\ket{\phi_1},\ket{\phi_2},\ket{\phi_3})$ be
$A_3$-invariant but not $S_3$-invariant and not zero.
By Lemma~\ref{antrivisn}, we must have $t_{(123)}\neq 1$. Without loss
of generality, suppose that $t_{(123)}=\omega^2=e^{4\pi i/3}$.
If
$\ket{\phi_1}=\ket{\phi_2}=\ket{\phi_3}$ (or possibly differing by a
global phase factor) then we have
$$\ket{\psi}=\GSym_t(\ket{\phi_1},\ket{\phi_2},\ket{\phi_3})\propto
(1+\omega+\omega^2)\ket{\phi_1}^{\otimes 3}=0.
$$
Conversely, suppose that $\ket{\psi}=0$.
Let
\begin{align*}
  \ket{\phi_1} &= a\ket{0}+b\ket{1}\\
  \ket{\phi_2} &= c\ket{0}+d\ket{1}\\
  \ket{\phi_3} &= e\ket{0}+f\ket{1}
\end{align*}
so that we have
\begin{align*}
 \ket{\psi}&= \GSym_t(a\ket{0}+b\ket{1},c\ket{0}+d\ket{1},e\ket{0}+f\ket{1})\\
&\propto \ket{M_3(bce+\omega acf + \omega^2 ade,bde+\omega bcf + \omega^2
  adf)}.
\end{align*}
Choosing a unitary operator $U$ so that $U\ket{\phi_1}=\ket{0}$ allows
us to set $a=1$ and $b=0$ in the above expressions, and we may work with the LU
equivalent state $\ket{\psi'}=U^{\otimes 3}\ket{\psi}$ of the form
\begin{align*}
 \ket{\psi'}&= \GSym_t(\ket{0},c\ket{0}+d\ket{1},e\ket{0}+f\ket{1})\\
&\propto \ket{M_3(cf + \omega de,\omega  df)}.
\end{align*}
The assumption that $\ket{\psi'}=0$ implies that $df=0$. If $d=0$, then
we have $\ket{\psi'}\propto \ket{M_3(cf,0)}$. Since $c\neq 0$
(otherwise, $\ket{\phi_2}=c\ket{0}+d\ket{1}=0$ would not be a state), so
we have $f=0$. This implies
$\ket{\phi_1}=\ket{\phi_2}=\ket{\phi_3}=\ket{0}$ up to phase. A similar
argument works for the case $f=0$. This completes the proof.

{\bf Proof of Proposition~\ref{A4symprop}.}

Let $G=A_4$ and let $\ket{\psi}=\GSym_t(\ket{\phi_1},\ket{\phi_2},\ket{\phi_3},\ket{\phi_4})$ be
$A_4$-invariant but not $S_4$-invariant and not zero.
By Lemma~\ref{antrivisn}, we must have $t_{(123)}\neq 1$. Without loss
of generality, suppose that $t_{(123)}=\omega=e^{2\pi i/3}$. By
Proposition~\ref{a4twoluclasses}, we know that $\ket{\psi}=\ket{M_4}$ or
its conjugate, or zero. By the
$U(2)^{\otimes 4}$ invariance of $\ket{M_4}$ and its conjugate~\cite{maxstabnonprod2}
, we may choose a unitary operator $U$ such that
$U\ket{\phi_1}=\ket{0}$, so that we have
\begin{align*}
  \ket{\psi} &= U^{\otimes 4}\ket{\psi}\\
  &=\GSym_t(U\ket{\phi_1},U\ket{\phi_2},U\ket{\phi_3},U\ket{\phi_4})\\
&=  \GSym_t(\ket{0},a\ket{0}+b\ket{1},c\ket{0}+d\ket{1},e\ket{0}+f\ket{1}).
\end{align*}
A straightforward calculation yields
\begin{equation}\label{a4symcalc}
  \ket{\psi} \propto (adf + \omega bde + \omega^2 bcf)\ket{M_4}.
\end{equation}
If three of the $\ket{\phi_i}$ are equal up to phase, say (without loss
of generality), $\ket{\phi_1}=\ket{\phi_2}=\ket{\phi_3}$, then we have
  $b=d=0$. Thus by~(\ref{a4symcalc}), we have
$\ket{\psi}=0$. Conversely, if $\ket{\psi}=0$, then we must have
$$adf + \omega bde + \omega^2 bcf=0.
$$
The solutions to this polynomial in the six state coefficients
$a,b,c,d,e,f$ is a set of measure zero in $\C^6$.

{\bf Proof of Proposition~\ref{spevencpodd}.} Suppose $J$ is of type \SP. Let $K$ be a self-palindromic
string in $[J]$, that is, such that $\tau K=K$, and choose $\ell$ so
that $\epsilon^\ell K = J$. Then we have
\begin{align*}
  \tau \epsilon^k J &= J\\
  \tau \epsilon^k \epsilon^\ell K &= \epsilon^\ell  K \hspace*{.2in}\mbox{(substituting $J = \epsilon^\ell K$)}\\
  \tau \epsilon^{k+\ell} \tau K &= \epsilon^\ell
  K \hspace*{.2in}\mbox{(using $\tau K =K$)}\\
   \epsilon^{-k-\ell}  K &= \epsilon^\ell K \hspace*{.2in}\mbox{(using
     $\tau \epsilon \tau = \epsilon^{-1}$)}\\
      \epsilon^{-k-2\ell}  K &= K
\end{align*}
We must have $m|(k+2\ell)$, so
$k$ is even. Now suppose $J$ is of type \CP. Choose $K\in [J]$ such that
$\tau K=\epsilon K$. A similar derivation to the one above leads to
$$      \epsilon^{-k-2\ell-1}  K = K
$$                                                                                                  so that $m|(k+2\ell + 1)$, and therefore $k$ must be odd.      

\begin{figure}
  \centering{
  \includegraphics[scale=.5]{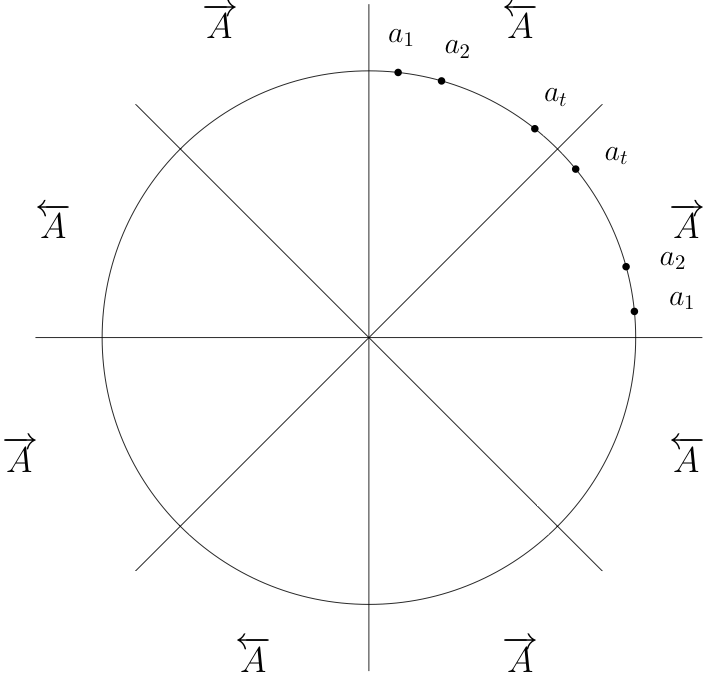}
  \caption{The cycle order of a necklace diagram is determined by the
    number of lines of mirror symmetry.}
  \label{cycle_order_fig}
  }
\end{figure}

{\bf Proof sketch for Proposition~\ref{cycleorderbymirrorlinect}.} 
In place of a formal proof, we illustrate the case for $L=4$. The lines
of symmetry partition the set of vertices into $2L$ subsets of equal size,
say $t=\frac{n}{2L}$, as in Figure~\ref{cycle_order_fig}. Let
$\overrightarrow{A}$ denote the bit string
$\overrightarrow{A}=(a_1a_2\cdots a_{t-1}a_t)$ in one of the regions between
the lines of symmetry, and let $\overleftarrow{A}=(a_ta_{t-1}\cdots
a_2a_1)$ be the reversed bit string. By reflection symmetry, the class
of the necklace diagram is $[\underbrace{\overrightarrow{A}\overleftarrow{A}\cdots
  \overrightarrow{A}\overleftarrow{A}]}_{L \text{ pairs}}$, and it is clear that the cycle
order of the necklace diagram is $2t = n/L$.

{\bf Proof of Theorem~\ref{cnnotdninv}.} We begin with the ``only if'' direction, that is, we
suppose $\ket{\psi}$ is $D_n$-invariant, and we will show that
properties (i)--(iv) must hold. Note that if $\ket{\psi}$ is $D_n$-invariant with
$\sigma\ket{\psi}=s_\sigma \ket{\psi}$ for all $\sigma\in D_n$, then we must
have $s_\epsilon=t_\epsilon$ and
$s_\tau=\pm 1$.
\begin{enumerate}
\item [(i)] The equation $\tau\epsilon\tau = \epsilon^{-1}$ in $D_n$
  implies that $s_\tau s_\epsilon s_\tau = s_\epsilon^{-1}$, so we have
  $s_\epsilon^2=s_\tau^2=1$, so $s_\epsilon=t_\epsilon=\pm 1$.
\item [(ii)] Suppose there is an $I$ of type \SP\ such that $c_{[I]}\neq
  0$. There is a string $J\in[I]$ such that $\tau J=J$, so we must have
  $s_\tau = 1$.
\item [(iii)] Suppose there is an $I$ of type \CP\ such that $c_{[I]}\neq
  0$. There is a string $J\in[I]$ such that $\tau J=\epsilon J$, so we must have
  $s_\tau = s_\epsilon$.
\item [(iv)] The equation $c_{J}=s_\tau c_{\tau J}$ 
  must hold for all $J$ by~(\ref{gitoJconstt}). In particular, it must hold for any such $J$ of
  chiral type.
\end{enumerate}

Now we prove the ``if'' direction. Suppose that conditions (i)--(iv)
hold. We will show that $\ket{\psi}$ is $D_n$-invariant by showing that
$\ket{\psi}$ has the Dicke form
\begin{equation}\label{dndickeform}
   \ket{\psi} = \sum_{[I]} d_{[I]} \sum_{J\in
  [I]}s_g^{-1}\ket{J} \hspace*{.2in}\mbox{ (where $gL_{[I]}=J$)}
\end{equation}
for a homomorphism $s\colon D_n\to U(1)$ with the property that $s$ is
constant on $D_n$-orbits $[I]$ for which $d_{[I]}\neq 0$. Note that
$[I]$ denotes the $D_n$-orbit of $I$ in~(\ref{dndickeform}). For $I$ of
type \SP\ or \CP, the $D_n$-orbit and $C_n$-orbit of $I$ are the
same. For $I$ of chiral type, the $D_n$-orbit of $I$ is the union
\begin{equation}\label{chiraldnunion2cns}
   [I]_{D_n} = [I]_{C_n} \cup [\tau I_{C_n}]
\end{equation}
of disjoint $C_n$ orbits. 

To show that $\ket{\psi}$ can be written in
the form~(\ref{dndickeform}), we start with $\ket{\psi}$ in $C_n$ Dicke
form. 
\begin{align*}
  \ket{\psi} &= \sum_{[I]} c_{[I]}
  \sum_{k=0}^{n-1}t_\epsilon^{-k}\ket{\epsilon^k L_{[I]}}\\
  &= \sum_{[I] \mbox{\tiny \SP type}}   \sum_{k=0}^{n-1}t_\epsilon^{-k}\ket{\epsilon^k L_{[I]}}\\
&+  \sum_{[I] \mbox{\tiny \CP type}}  \sum_{k=0}^{n-1}t_\epsilon^{-k}\ket{\epsilon^k L_{[I]}}\\
  &+  \sum_{[I]_{D_n} \mbox{\tiny chiral type}}\\
  &\left(c_{[I]}
  \sum_{k=0}^{n-1}t_\epsilon^{-k}\ket{\epsilon^k L_{[I]}}
  + c_{[\tau I]} \sum_{k=0}^{n-1} t_\tau^{-1}t_\epsilon^{-k}
  \ket{\tau\epsilon^k \tau L_{[I]}} \right)\\
  &\hspace*{.2in}\mbox{(using~(\ref{chiraldnunion2cns}))}\\
  &= \sum_{[I] \mbox{\tiny \SP type}}   \sum_{k=0}^{n-1}t_\epsilon^{-k}\ket{\epsilon^k L_{[I]}}\\
&+  \sum_{[I] \mbox{\tiny \CP type}}  \sum_{k=0}^{n-1}t_\epsilon^{-k}\ket{\epsilon^k L_{[I]}}\\
 &+  \sum_{[I]_{D_n} \mbox{\tiny chiral type}} c_{[I]}
 \sum_{k=0}^{n-1} t_\epsilon^{-k}\left(\ket{\epsilon^k L_{[I]}}
 +  \ket{\tau\epsilon^k \tau L_{[I]}} \right)\\
 &\hspace*{.2in}\mbox{(using (iv))}
\end{align*}
By Proposition~(\ref{dickeformisginv}), all that remains to be shown is that $s_\tau$ can be assigned in a way
that guarantees the following condition.

\begin{equation}\label{dncompatiblecond}
\mbox{if $\tau \epsilon^k J = J$ for some $c_{[J]}\neq
0$, then $s_\tau s_\epsilon^k = 1$} 
\end{equation}
Because $\tau\epsilon^k$ cannot stabilize any $J$ of chiral type, it
suffices to show that~(\ref{dncompatiblecond}) holds for $J$ of type
\SP\ or \CP. We consider cases.

Suppose there is an $I$ of type \SP\ with
$c_{[I]}\neq 0$ and a $J$ of type \CP\ with $c_{[J]}\neq 0$. Then by~(ii)
and~(iii), we have $s_\tau=s_\epsilon = 1$, so~(\ref{dncompatiblecond})
holds. 

Suppose there is an $I$ of type \SP\ with $c_{[I]}\neq 0$, but there is
no $J$ of type \CP\ with $c_{[J]}\neq 0$. By~(ii), we have
$s_\tau=1$. If the cycle order of $I$ is odd, then $s_\epsilon=1$
($s_\epsilon$ is a power of $e^{e\pi i/m}$ for an odd number $m$, so
$s_\epsilon$ cannot equal minus 1),
so~(\ref{dncompatiblecond}) holds. If the cycle order of $I$ is even, then by
Proposition~\ref{spevencpodd}, we have that $k$ must be even if $\tau\epsilon^k I = I$, so~(\ref{dncompatiblecond})
holds.

Finally, suppose there is an $I$ of type \CP\ with $c_{[I]}\neq 0$, but
there is no $J$ of type \SP\ with $c_{[J]}\neq 0$. By~(iii), we have
$s_\tau=s_\epsilon$. If the cycle order of $J$ is odd, we have
$s_\epsilon = 1$.  If the cycle order of $I$ is odd, then
$s_\epsilon=1$, so~(\ref{dncompatiblecond}) holds. If the cycle order of
$I$ is even, then by Proposition~\ref{spevencpodd}, we have that $k$
must be odd if $\tau\epsilon^k I = I$, so~(\ref{dncompatiblecond})
holds.

This concludes the proof of the Theorem.

{\bf Proof of Proposition~\ref{dnnotsninv}.} We begin with the ``only
if'' direction, that is, we suppose $\ket{\psi}$ is $S_n$-invariant with
$\sigma \ket{\psi} = s_\sigma \ket{\psi}$ for all $\sigma \in
S_n$. Proposition~\ref{singletonlyphaseinv} states
that if $\ket{\psi}$ is $S_n$-invariant where $n\geq3$, then
$t_\sigma=1$ for all $\sigma \in S_n$. So, we know that
$t_\epsilon=t_\tau=1$. Since $\ket{\psi}$ is $S_n$-invariant, all terms
of the same weight must share the same coefficient, otherwise a series
of transpositions would be unable to take one term to the other while
also accounting for this coefficient shift. Thus, we can conclude
$C_{[I]}=C_{[J]}$ for all $I,J$ with $\wt I= \wt J$.

Now looking at the ``if'' direction, we suppose that the two statements
hold to prove that $\ket{\psi}$ is $S_n$-invariant. Let $C_w$ be the
common value of $C_{[I]}$ for all $I$ with weight $w$, then
\begin{align*}
    \ket{\psi} &= \sum_{[I]}C_{[I]}\sum_{J\in [I]}t_g^{-1} \ket{J}\\
    &= \sum_{[I]}C_{[I]}\sum_{J\in [I]} \ket{J} \hspace*{.2in}\mbox{(using (i))}\\
    &= \sum_{w=0}^{n}C_w\sum_{\wt J = w} \ket{J}
\end{align*}
Evidently, this is the Dicke form for an $S_n$-invariant state, thus completing the proof.

{\bf Proof of Proposition~\ref{weightsludistinct}.}
The proof is the same as the proof for Theorem~1
in~\cite{symmstatespaper}, with the observation that full permutational
symmetry may be replaced by cyclic symmetry in places where symmetry is
needed. We will not reproduce the full proof here, which requires
lengthy technical preliminaries. Instead we provide a sketch of the main
ideas.

Let $\ket{D^w}$ denote a generalized Dicke state $\ket{D_t^{[I]}}$ for
some bit string $I$ with $\wt I=w$, for $G=S_n$, $G=C_n$, or $G=D_n$. It
does not matter whether the phase homomorphism $t\colon G\to U(1)$ is
trivial, and it does not matter what particular $G$-orbit class is for
$I$. Every part of the proof depends only on the weight of $I$, and on
the fact that $D^w$ has at least cyclic symmetry.

From the observation that
$$\left(e^{itZ}\right)^{\otimes n}\ket{D^w} = e^{it(n-2w)}\ket{D^w}
$$
we have that the group
\begin{equation}\label{stabdw}
  \left\{e^{it(2w-n)} \left((e^{itZ}\right)^{\otimes n}\colon t\in \R\right\}
\end{equation}
is contained in the local unitary stabilizer $\Stab_{D^w}$. We view the local
unitary group as the Lie group $U(1)\times SU(2)^n$, with Lie algebra
$u(1)\oplus \bigoplus_{j=1}^n su(2)$, where $u(1)$ is the real vector space
$u(1)=\{it\colon t\in \R\}$ and $su(2)$ is the real vector space of
traceless skew-Hermitian matrices. 
The Lie algebra of~(\ref{stabdw}) is the real vector space
\begin{equation}\label{lstabdw}
   \left\{it[(2w-n) + Z^{(1)} + Z^{(2)} + \cdots +Z^{(n)}] \colon t\in \R\right\}
\end{equation}
  so that~(\ref{lstabdw}) is a subspace of the Lie algebra of the group
  $\Stab_{D^w}$. The notation $Z^{(k)}$ denotes the Pauli $Z$ operator
  acting on the $k$-th qubit. The heart of the proof is an argument that
  in fact,~(\ref{lstabdw}) is the entire Lie algebra of $\Stab_{D^w}$. The idea is
  that if there were some stabilizer Lie algebra element $is +
  \sum_{k=1}^n M_k^{(k)}$ with $M_k$ independent of $Z$, then there
  would also be an element in the stabilizer Lie algebra with $[M_k,Z]$
  in the $k$th summand, and therefore the projection of the stabilizer
  Lie algebra in the $k$-th position would be three dimensional. By the
  cyclic symmetry of $D^w$, it would follow that the projection of the
  stabilizer Lie algebra would be three dimensional in all qubits. But
  we have classified all possible states whose Lie algebra stabilizers
  have three dimensional projections in each qubit (these states are
  superpositions of products of singlet states,
  see~\cite{su2blockstates}), and these states are {\em not} local
  unitary equivalent to $D^w$. Thus, it must be that~(\ref{lstabdw}) is all of the
  Lie algebra of $\Stab_{D^w}$, and therefore, that~(\ref{stabdw}) is
  all of the connected component of $\Stab_{D^w}$ that contains the
  identity.

  Now suppose that there is some local unitary operator $U=U_1\otimes
  U_2\otimes \cdots \otimes U_n$, for some $2\times 2$ unitary operators
  $U_1,U_2,\ldots, U_n$, that takes $\ket{D^w}$ to $\ket{D^{w'}}$. It
  follows that $\Stab_{D^{w'}} = U\Stab_{D^w}U^\dagger$. For each $k$,
  we must have $U_k Z U_k^\dagger \propto Z$. It is a simple exercise to
  se that this implies $U_k = \pm I,\pm X$, and therefore
  $U_kZU_k^\dagger = \pm Z$. It turns out that all of the $U_k$ must be
  equal, so we have
  \begin{align}\nonumber
    &U(it(2w-n) + Z^{(1)} + \cdots + Z^{(n)})U^\dagger\\
    &= it(2w-n) \pm (Z^{(1)} + \cdots + Z^{(n)})\label{pmzs}
  \end{align}
  In the case where the last expression~(\ref{pmzs}) is
  $$it(2w-n) + (Z^{(1)} + \cdots + Z^{(n)}),
  $$
  we conclude that $w'=w$, and in the case where the last
  expression~(\ref{pmzs}) is
    $$it(2w-n) - (Z^{(1)} + \cdots + Z^{(n)}),
  $$
  we conclude that $w'=n-w$. This completes the sketch of the proof.

\end{document}